\documentclass[twocolumn, prl]{revtex4}
\usepackage[utf8]{inputenc}
\usepackage{amsmath}
\usepackage{algorithm}
\usepackage{algorithmicx}
\usepackage{algpseudocode}
\usepackage{graphicx}
\usepackage{subfigure}
\usepackage{hyperref}

\begin{document}
\title{Subgraph covers-An information theoretic approach to motif analysis in networks }
\author{Anatol E. Wegner}
\affiliation{Max Planck Institute for Mathematics in the Sciences, Inselstr. 22, Leipzig-Germany}
\email{wegner@mis.mpg.de}

\begin{abstract}
Many real world networks contain a statistically surprising number of certain subgraphs, called network motifs. In the prevalent approach to motif analysis, network motifs are detected by comparing subgraph frequencies in the original network with a statistical null model. In this paper we propose an alternative approach to motif analysis where network motifs are defined to be connectivity patterns that occur in a subgraph cover that represents the network using minimal total information. A subgraph cover is defined to be a set of subgraphs such that every edge of the graph is contained in at least one of the subgraphs in the cover. Some recently introduced random graph models that can incorporate significant densities of motifs have natural formulations in terms of subgraph covers and the presented approach can be used to match networks with such models. To prove the practical value of our approach we also present a heuristic for the resulting NP-hard optimization problem and give results for several real world networks.
\end{abstract}
\maketitle
\section{Introduction}
Many complex systems can be modeled as networks where vertices represent interacting elements and edges interactions between them. A large number of real world networks has been found to contain a statistically surprising number of certain small connectivity patterns called network motifs \cite{milo}. Network motifs, which are also commonly referred to as basic building blocks of networks, are thought to play an important role in the structural and functional organization of complex networks. For instance, in biological and technological networks motifs are thought to contribute to the overall functioning of networks by performing modular tasks such as information processing \cite{alon}. Hence, methods for identifying such characteristic connectivity patterns are of great importance for a better understanding of complex networks.  

The prevalent approach to motif analysis is due to Milo et al.\cite{milo} and is based on comparing the subgraph frequencies of the original network with a statistical null model that preserves some features of the original network. Part of the analysis consists of generating a representative sample of the null model which is used to determine empirical values for the mean and variance of motif counts in the null model. Motifs for which the frequencies significantly deviate from the null model are then classified as network motifs. In their original paper,  Milo et al. suggest that when detecting motifs of size n the null model should conserve the degree distribution of the original network as well as the motif counts of size n-1. For generating such networks they propose a simulated annealing approach. However, it is not clear whether the simulated annealing approach samples such null models uniformly. Moreover, in most applications it is computationally not feasible to preserve lower order motif counts for motifs larger than 4 vertices. Consequently, in most practical applications \cite{milo, fanmod, mavisto} lower order motif counts are not conserved and the configuration model \cite{newmandeg} with the same degree distribution as the original network is used as a null model. This has the unwanted consequence that most subgraphs that contain an over-represented sub-motif are classified as network motifs.

In this paper, we introduce an alternative approach to motif analysis that is based on using subgraph covers as representations of graphs. A subgraph cover can be seen as a decomposition of the network into smaller building blocks. Given any network there are many ways of decomposing it into a subgraph cover. Consequently, one needs a way of comparing subgraph covers. For this, following the total information approach by Gell-Mann and Lloyd \cite{gellmann}, we look at motifs as regularities of a network which can be used to obtain a more concise representation of the network. In our approach network motifs are defined as subgraph patterns that appear in a subgraph cover that represents the network using minimal total information. Note that this definition of network motifs is fundamentally different from the definition of Milo et al. \cite{milo}. 

Another aim of this paper is to establish a connection between motif analysis and random graph models. In contrast to most real world networks, commonly used network models are locally tree-like. Developing random graph models that can incorporate high densities of triangles and other motifs has been a long standing problem.  Recently, two random graph models that can incorporate significant densities of motifs have been proposed \cite{bollobasc,karrer}. However, it remains unclear how one should select the set of motifs to be used in such models given a specific network. As we shall see later, these models can be formulated as ensembles of subgraph covers and total information optimal subgraph covers can be used to match networks with specific instances of these models.  

The article is organized as follows: in Sec.2 we present the theory underlying our approach. Then in Sec.3, we examine the resulting optimization problem and propose a heuristic for it. In Sec. 4 we present empirical results for several real world networks and also test the heuristic on some synthetic networks with predefined motif structure. Finally, in Sec.5 we summarize our results and discuss directions for future research. 

\section{Theory}
In this section we first introduce necessary graph and information theoretical concepts. We then define the total information for subgraph covers and following the approach by Gell-Mann and Lloyd  \cite{gellmann}, we use the smallness of the total information as a criterion for selecting a subgraph cover that is an optimal representation of a given network. Finally, we discuss the relation between total information optimal subgraph covers and model selection for random graphs.

\subsection{Subgraph Covers}

A graph $G=(V(G),E(G))$ is a ordered pair of sets where $V(G)$ ($|V(G)|=N$) is the set of points called vertices and $E(G)$ is a set of links called  edges that connect pairs of vertices. Depending on the kind of network, edges might be directed or undirected. Though in this article, we will not make an explicit distinction between directed and undirected graphs since the arguments and definitions apply to both equally well. In general, we will assume that $G$  is sparse, i.e. $|E(G)|=O(N)$. Most real world  networks are sparse \cite{newmanns}. 

In graph theory, motifs correspond to isomorphism classes. Two graphs $G$ and $H$ are said to be isomorphic ($G\simeq H$) whenever there exist a bijection $\phi:V(G)\rightarrow V(H)$ such that $(x,y)\in E(G) \Leftrightarrow (\phi(x),\phi(y))\in E(H)$ for all x,y $\in V(G)$. Such a map $\phi$ is called an isomorphism. Being isomorphic is an equivalence relation and the corresponding equivalence classes are called isomorphism classes.

A graph $H$ is called a subgraph of $G$  whenever $V(H)\subseteq V(G)$ and $E(H)\subseteq E(G)$. A set of subgraphs $C$, is said to be a subgraph cover of $G$   whenever $\bigcup_{H \in C}E(H)=E(G)$. Subgraph covers are representations of graphs meaning that given a subgraph cover the corresponding graph can be recovered fully from the cover. Trivial examples of subgraph covers are the set of all edges of $G$  and $G$  itself. Other examples are the maximal clique and star covers, which are the sets of all cliques/stars that are not subcliques/substars. An n-clique consists of n vertices that are all mutually connected and an n-star consists of single central vertex that is connected to n peripheral vertices and a subclique/substar is a clique/star that is a subgraph of some larger clique/star. While the maximal star cover is closely related to the adjacency list representation of the graph, clique covers are closely related to bipartite representations \cite{newmandeg,latapy}.

Given a cover $C$, its motif set $M(C)$ is the set of the isomorphism classes of the subgraphs in $C$. Similarly given a set of isomorphism classes $M$, an $M$-cover $C_M$ is a subgraph cover of which every element belongs to some class in $M$. 

\subsection{The Total Information Approach}

The total information framework \cite{gellmann} is based on the idea that, given an entity, one can use Shanon information or entropy to describe its random/non-regular aspects and algorithmic information content to describe its regularities or rule based features. In this approach, identifying certain regularities of an entity  is equivalent to embedding it into an ensemble of objects that share these regularities while they differ in other aspects. 

The first information measure of interest for the total information approach is the entropy, also known as the Shannon information. For an ensemble $E(R,p_r)$, where $R$ the is the set of possible outcomes and $p_r$ is the respective probability of an element $r\in R$, the entropy measures the uncertainty regarding the outcome of $E$ and is given by:
\begin{equation}
S=-K\sum_{r} p_{r} \log p_{r},
\end{equation}
where $K$ is a constant. When $K=1$ and logarithm is base 2, the entropy is measured in bits.   

Another information measure that is needed in order to define the total information is the effective complexity. The effective complexity ($\epsilon(E)$) of an entity that is embedded into an ensemble E as a typical member is given by the algorithmic information content (AIC)  of the ensemble. The algorithmic information content of such an ensemble E with respect to a universal computer U is the length of the shortest program that instructs U to output a description of E and then halt (i.e. $\epsilon(E)=AIC_U(E)$). In general the effective complexity is not computable and computer dependent, therefore in practice one is restricted to work with approximations in the form of upper bounds. The issue of how to define a practical effective complexity for subgraph covers is dealt with in Sec.3.

The sum of the effective complexity and the entropy is the total information required to describe both the random features and regularities of an entity using a certain model:
\begin{equation}
 \Sigma(E)=\epsilon(E) + S(E)
\end{equation}

For a given entity, there might be a multitude of ensembles into which the entity can be embedded as a typical member and it may not always be clear which set of regularities/model provides the best description of the entity. The total information provides a basis for comparing models that describe the same entity. When comparing models, the better model is the one that minimizes the total information and then subject to this constraint, minimizes the effective complexity. Together with additional constraints on computation time, the framework provides a method for identifying regularities/models that 'most' effectively describe a given entity which in many regards is independent of the observer \cite{gellmann}. The minimization of the total information is closely related to the minimum description length \cite{mdl} and minimum messaging length approaches \cite{mml}.  

\subsection{Uniform subgraph covers}
Following the above definitions, we define the total information of subgraph covers by embedding them into uniform subgraph covers. These are the ensembles of all subgraph covers with fixed motif counts. For this we need to compute the number of different ways a motif $m$ can appear on $N$  vertices which depends on the automorphism group of the motif. An automorphism of a graph is a permutation of its vertex labels that preserves the edges of the graph. The number of all such permutations gives us the number of equivalent vertex labellings of the graph. To specify an instance of $m$ on $N$  vertices, one needs to specify the set of vertices $m$ appears on and how it is embedded into this set. From the definition of the automorphism group it follows that there are $\frac{|m|!}{|Aut(m)|}$ different ways a motif $m$ can appear on a set of $|m|$ vertices.  Thus a motif $m$ with automorphism group $Aut(m)$ can appear on $N$  vertices in $\frac{N!}{(N-|m|)!|Aut(m)|}$ different ways. Consequently, the entropy of a set of $n_m$ distinct instances of $m$ is given by:
\begin{equation}
S(m,n_{m})=\log \left( \begin{array}{c} 
\frac{N!}{(N-|m|)!|Aut(m)|} \\
n_m\end{array}\right),
\end{equation}
which, given $m$, is the information required to specify $n_m$ instances of $m$ on $N$  vertices. 

Generalizing the above expression, the entropy of a cover $C$ with motif set $M(C)$ and motif counts $n_m$ ($m\in M$) is defined as the entropy of the uniform ensemble of all covers with motif counts $n_m$:
\begin{equation}
S(C)=\sum_{m \in M(C)}S(m,n_m).
\end{equation}

When needed, the entropy terms can easily be approximated using Stirling's formula. For instance, when $n_m$ and $N$  are large enough and $|m|>2$:
\begin{widetext}
\begin{equation}
 S(m,n_m)= n_m(|m|\log N-\log|Aut(m)|-\log n_m+\log(e))+O(\log N). 
\end{equation}
\end{widetext}

As in the case of the entropy, we define the effective complexity of a cover using uniform covers with the same motif counts: $\epsilon(C)=AIC_U(E(M(C),n_m))$. Consequently, the total information of a cover can be defined as: 
\begin{equation}
\Sigma(C)= \epsilon(C)+S(C).
\end{equation}

Following the total information approach, a cover is an optimal representation of the network if it minimizes the total information. As a result we can define the network motifs of $G$  to be the motif set of a $\Sigma$-optimal subgraph cover of $G$: $M(C_{\Sigma}(G))$.  The $\Sigma$-optimal subgraph cover also gives a decomposition of the network in terms of these motifs. In general there might be more than one subgraph cover that minimize $\Sigma$. If this is the case, additional criteria such as the minimization of the effective complexity \cite{gellmann} have to be considered in order to pick one of the solutions over the others.  

The quantity $\epsilon(C_{\Sigma}(G))$ can be interpreted as a measure of the complexity of $G$'s subgraph structure which actually is in correspondence with other measures that are frequently used as indicators of a network's complexity such as the broadness of the degree distribution and/or clustering. While the broadness of the degree distribution gives the variety of the star shaped subgraphs that occur in the network, high clustering indicates that the network has a local structure that involves subgraphs other than trees. 

Another quantity of interest is the amount of compression the optimal cover provides with respect to the edge cover: 
\begin{equation}
 \Delta_\Sigma(G)=\Sigma(C_e(G))-\Sigma(C_{\Sigma}(G)).                                                                                                                     
\end{equation}
This gives us a measure of how much $G$  deviates from a Erdös-Rényi type random graph with respect to its subgraph structure. Motifs can also be assigned a quantitative significance similar to the z-score used by Milo et al. \cite{milo} based on the compression they provide. One such measure is the c-score $c_m$: 
\begin{equation}
 c_m(G)=\frac{\Sigma(C_{\Sigma}(G)- m)}{\Sigma(C_{\Sigma}(G))}-1,
\end{equation}
where $(C_{\Sigma}(G)-m)$ is the cover obtained by replacing the $m$-subgraphs in $C_{\Sigma}(G)$ with the single edge subgraphs they cover.  By definition $c_m$ is always non-negative and zero for the single edge motif and motifs that are not contained in $C_{\Sigma}(G)$.

\subsection{Subgraph Covers and Model Selection}

In this section we will consider two models that are closely related to subgraph covers: the model introduced by Bollobàs, Janson and Riordan \cite{bollobasc} and the model introduced by Karrer and Newman \cite{karrer}. Although these models can account for large densities of nontrivial subgraphs, it is not clear how one should select the set of motifs to be used in such models when matching these with a given network. In their article Karrer and Newman \cite{karrer} mention this as an important open problem. In the following, we formulate these models in terms of subgraph covers and discuss how $\Sigma$-optimal subgraph covers can be used to associate networks with such models. 

\textbf{Random graphs with clustering:} In \cite{bollobasc} Bollobàs, Janson and Riordan introduced a very general class of random graph models that is based on adding copies of certain motifs on to the vertices of a graph. For the sake of simplicity, we will only consider the homogeneous models i.e. the case where all vertices have the same type. For the non-homogeneous version of the model as well as various analytical results concerning the properties of the model we refer the reader to the original paper \cite{bollobasc}.  In the homogeneous case the model can be defined as follows: Let $M$ be a set of motifs, each given by a labeled representative and for every $m\in M$, let $k_m$ be positive constant that corresponds to the density of the motif in the model. Then for each $m\in M$ and $|m|$-tuple $(v_1,v_2,...,v_{|m|})$ of vertices one adds a copy of $m$ to $G$, such that $i^{th}$ vertex of $m$ is mapped onto $v_i$, with probability:
\begin{equation}
 p_m=\frac{k_m}{N^{|m|-1}}.
\end{equation}
Since we are mainly interested in simple graphs, we will assume that any parallel edges that are formed in this process are replaced with single edges in the network. The normalization factor $1/N^{|m|-1}$ ensures that the model has O(N) edges. Depending on the symmetry of $m$, the same subgraph might be added to the graph more than once, although the probability of this is very small. However, the model can be slightly modified in such a way that every $m$-subgraph is only considered once. This can be done by considering $|m|$-subsets of vertices instead of $|m|$-tuples. Then for each such subset, every distinct $m$-subgraph is added with probability: 
\begin{equation}
 p_m=\frac{k_m|Aut(m)|}{N^{|m|-1}},
\end{equation}
where the factor $|Aut(m)|$ ensures that both models contain the same number of copies of $m$ on average. 

With the above modification the model defines a multinomial distribution $P_{(M,\textbf{k})}(\cdot)$ over the space of $M$-covers. This is then projected on to the set of edges in order to obtain a distribution over graphs. Thus, the probability of a graph $G$  in this model is given by: 
\begin{equation}
 P_{(M,\textbf{k})}(G)=\sum_{C \in \mathcal{C}_M(G)} P_{(M,\textbf{k})}(C),
\end{equation}
where $\mathcal{C}_M(G)$ is the set of all $M$-covers of $G$.

Uniform subgraph covers are essentially micro-canonical versions of these models. Consequently, the presented approach can be seen as way of inferring the subgraph cover state of such models. The $\Sigma$-optimal cover can be further used as a basis for associating the network with non-homogeneous models of this type that also include correlations between subgraphs.  

\textbf{Generalized configuration models:} Another random graph model that is closely related to subgraph covers is the generalized configuration model proposed by Karrer and Newman \cite{karrer}. This model is defined on the basis of a motif set $M$ and a corresponding role sequence \textbf{r}. Here, the role sequence specifies the number of different motifs attached to each vertex and how these motifs are attached to the vertex. The way in which a certain motif is attached to the vertex is given by the orbit of the automorphism group of the motifs the vertex belongs to. The orbit of a vertex is the set of vertices it can be mapped onto by the automorphism group. In order to generate a graph corresponding to a role sequence \textbf{r}, every vertex is assigned a number of subgraph-stubs corresponding to its role index. A graph is then generated by matching stubs corresponding to the same type of subgraph in appropriate combinations at random and connecting them in order to form the corresponding motif-subgraph.  However, in this form the model allows for two or more stubs of the same vertex to be matched together which results in a subgraph that is a vertex contraction of the original motif. When such problematic cases are excluded from the model, every matching of the stubs actually corresponds to an M-cover. Consequently, the generalized configuration models can be formulated in terms of subgraph covers: the model corresponding to a role sequence \textbf{r} is the uniform ensemble of all subgraph covers with role sequence \textbf{r}. 

Determining a role sequence for a network is essentially equivalent to choosing a subgraph cover for the network since every subgraph cover produces a specific role sequence for the network. The $\Sigma$-optimal cover can be considered as a viable candidate for assigning a role sequence to a network. On the other hand, an important property of the generalized configuration models is that biconnected subgraph counts are essentially determined by the motif set while singly connected subgraphs can be mostly accounted for by the role sequence. Consequently, one can also consider restricting the analysis to biconnected subgraphs when determining a role sequence for the network. This also significantly reduces the number of subgraphs that have to be considered in the analysis since the majority of connected subgraphs of sparse networks are only singly connected. 

The models described above suggest that in principle one could also consider more general/non-uniform ensembles of subgraph covers to define the total information. For instance, in the case of the generalized configuration model one could use the ensemble of all subgraph covers that result in the same role sequence. However, there is no known simple way of calculating the entropy of such ensembles even if only single-edge subgraphs are considered, which would be equivalent to the classical configuration model. In addition, such ensembles have high effective complexity.  

\subsection{The relation to the method of Milo et al.}
From a compression point of view the motifs of the $\Sigma$-optimal cover correspond to the motifs with respect to which the graph maximally deviates from a random graph. Although the Erdös-Rènyi random graph is in general not the null model of choice in applications of the method of Milo et al., both methods can be seen as trying to find motifs with respect to which the network differs from a random graph. In this sense the subgraph cover approach and the method of Milo et al. share similar goals. Using generalized configuration models to define the total information would allow for a more direct comparison of the two approaches but as explained in the previous section there are additional difficulties associated to this. 

A general problem one faces when using the method of Milo et al. is that subgraph frequencies are in general highly interdependent. The most obvious of such dependencies are motif-submotif type dependencies. That is, the presence of a motif implies the presence of larger motifs containing it as a submotif and its own submotifs. This in turn implies that counts of motifs that have a submotif in common are correlated. In order to avoid larger motifs to be classified as network motifs only because they contain some smaller overrepresented motif, Milo et al. propose using a null model that conserves lower order motifs. However, this does not cover all dependencies. For instance, a certain overrepresented motif might occur almost exclusively as a submotif of one or more larger network motifs. The $\Sigma$-optimal cover naturally avoids such interdependencies by considering motifs of all sizes simultaneously and effectively penalizing the sharing of edges between subgraphs. Despite their differences, one would expect both methods to find similar motifs provided that lower order motifs are conserved by the null model. 

\section{The $\Sigma$-optimal subgraph cover problem}

In general when finding optimal subgraph covers, one would like to consider the most general set of potential motifs. However, in practice there are several technical limitations, the first being the graph isomorphism problem. That is, there exists no known polynomial time algorithm for resolving the problem of whether two finite graphs are isomorphic. The same holds for finding the automorphism group of a graph. Fortunately, there are several software packages that can efficiently compute the automorphism group of small graphs \cite{nauty}. Second, the problem of finding whether a graph $G$  contains a certain motif as a subgraph is NP-complete. Thus, finding subgraph instances can be computationally expensive especially for large motifs. Third, the number of connected motifs grows faster than exponentially with size. For instance, there are over a million different directed motifs of size 6. Therefore, the set of candidate motifs of which the subgraph instances are to be included in the analysis has to be restricted so that the analysis can be completed in reasonable time.  Restricting the set of candidate motifs to all connected motifs up to a certain size seems to be an obvious choice. On the other hand, one can also include special classes of motifs of arbitrary size into the set of candidate motifs. 

If one wants to include special classes of motifs into the set of candidate motifs, any prior knowledge of the structure of the network can be used to make an educated guess about which motifs are more likely to produce covers with small total information. For instance, when examining the network representing an electronic circuit, the motifs corresponding to various known subcomponents of the circuit should be included in the set of candidate motifs. Also if the network at hand is known to have a broad degree distribution, star shaped motifs can be included. Similarly, if some motifs are known to favor a certain type of dynamical behavior that is thought to be relevant to the network performing certain tasks, these patterns and their generalizations can be included into the analysis. As previously mentioned, if one intends to use the subgraph cover in order to determine a role degree sequence for the network, the set of motifs can be restricted to biconnected motifs. Disconnected motifs can be excluded from the analysis since the cover that independently contains the connected components of such subgraphs always has lower total information. 

Another issue that has to be addressed in practice is that the algorithmic information content is not computable and in addition it is computer dependent. This can be resolved by substituting the algorithmic information content of the ensemble with the code length of a reasonable encoding of it. Another simplification we make is to assume that motifs are independently encoded which results in a effective complexity term that is additive in the motifs. One obvious way of encoding motifs is to use edge lists. In this case we have:
$$\epsilon(m)=\log^*(|V(m)|)+\log^*(|E(m)|)+S(|V(m)|,|E(m)|),$$
where $S(|V(m)|,|E(m)|)$ is the entropy of the ensemble of all graphs with the same vertex and edge counts as $m$ and $log^*$ is the iterated logarithm. On the other hand, one can also use a predefined/fixed  encoding or catalog of the candidate motifs to define their effective complexities. 

After the simplifications above, the total information reduces to:
\begin{equation}
\Sigma(C)=\log^*N + \sum_{m\in M(C)}\left(S(m,n_m)+\epsilon(m)+log^*n_m\right).
\end{equation}

The choice of code used to define the effective complexity depends on the set of candidate motifs. The edge list encoding has the advantage of being independent of the set of candidate motifs and therefore is a natural choice when considering all motifs up to a certain size. On the the hand, given a specific set of candidate motifs, the catalog approach in general results in shorter code lengths compared to the edge list encoding. This makes the catalog approach more suitable when the set of candidate motifs contains special classes such as cliques, stars, cycles etc. since these have obvious better/shorter encodings than their edge lists. 

Although the edge list encoding seems to be a natural candidate when considering general motifs, it might be argued that the choice of code used to approximate the effective complexity term is subjective. However, in practice almost all comparable model selection approaches require similar subjective choices \cite{mdl,mml,mckay}. From a Bayesian viewpoint, where the effective complexity term can be associated to a prior, this is less of a problem. The main goal of this article is not to advocate a specific approach to model selection but rather to show that subgraph covers can be used as a basis for motif analysis. The reason for us choosing the total information approach is that it accounts for the fact that the parameters of the ensembles are graphs/motifs in a rather intuitive way. 

Another, more practical, way of looking at the effective complexity is as a safeguard against overfitting. From this point of view the effective complexity of a motif corresponds to the minimal entropy gain it has to provide in order to be included in the optimal cover. This in turn can be seen as setting a frequency threshold for the motif. Thus, the problem can also be formulated in terms of entropy minimization and frequency constraints. If overfitting is less of a concern, one can even set the effective complexity to zero which in our case is equivalent to the maximum likelihood approach. Using maximum likelihood or similar reduced effective complexity terms/frequency thresholds might be useful if one wants to find a maximal number of potentially relevant motifs or when the network is small $(N<200)$ since in such networks the maximal potential entropy gains of motifs are limited. The algorithms we shall present in the next section can be modified in a straightforward manner to incorporate frequency thresholds.    

\subsection{The greedy algorithm}
Even with the candidate motifs restricted, finding a $\Sigma$-optimal subgraph cover is a non-trivial optimization problem. As formulated above, the problem of finding a $\Sigma$-optimal subgraph cover is a nonlinear set covering problem where the set to be covered is the edge set of the graph and the subsets are the edge sets of the subgraph instances of the candidate motifs. Set covering problems are known to be NP-hard even in the linear case \cite{NP}. Consequently, in most practical applications exact solutions are elusive and a heuristic has to be used. 

The greedy algorithm we propose might be seen as a generalization of the greedy heuristic introduced by Chvatal \cite{chvatal} for the linear set covering problem. The algorithm is based on the stepwise construction of a subgraph cover. At each step the algorithm finds the motif that covers not yet covered edges of $G$  most efficiently in terms of total information per edge. Given a partial cover $C$, the efficiency of a set $S_m$ of $m$-subgraphs is defined as:
\begin{equation}
 \sigma(S_m,C)=\frac{\Sigma(S_m)}{|E(S_m)-E(C)|},
\end{equation}
where $E(C)$ and $E(S_m)$ are the set of edges covered by $C$ and $S_m$ respectively and $\Sigma(S_m)$ is the total information corresponding to $S_m$. More precisely, $\Sigma(S_m)=S(m,|S_m|)+\epsilon(m)+log^*(|S_m|)$. Following this definition, an optimal instance set of $m$ is defined as a set of $m$- subgraphs that minimizes $\sigma$. At each step, the algorithm determines the efficiency of all motifs in the candidate motif set by determining an optimal instance set for each of them. In the next step, the algorithm checks for each motif whether including its optimal instance set into the cover increases the overall total information of the cover. Then from the motifs of which the optimal instance set does not increase the total information, most efficient one is selected. Having found the most efficient motif, the corresponding optimal instance set is added to the cover and the set of covered edges is updated. The process is repeated until all edges of the graph are covered. To ensure that the algorithm terminates, we require that the single edge motif is always included in the set of candidate motifs. The total information of partial covers is calculated by adding to them the single edge subgraphs corresponding to the uncovered edges. Here, one should note that motifs can not be selected based solely on their efficiency  because in general adding the optimal instance set of a motif to the cover decreases the efficiency of other motifs which, in certain cases might lead to an increase of the overall total information.   

\begin{algorithm}[H]

\begin{algorithmic}
\State $CoveredEdges=\emptyset,Cover=\emptyset,Motifs=\emptyset$ 
\While {$|CoveredEdges|<|E|$}
\State $C,m=FINDMOTIF(G,MS,CoveredEdges)$
\State $CoveredEdges\leftarrow CoveredEdges\cup_{i \in C} e(i)$
\State $Cover \leftarrow Cover\cup C$
\State $Motifs\leftarrow Motifs\cup \{m\}$
\EndWhile
\State \Return $Cover$, $Motifs$
\Function{FINDMOTIF}{$G,MS,CoveredEdges$}
\For {$m \in MS$}
\State $C(m)=OptimalInstanceSet(m,CoveredEdges,G(E,V))$
\EndFor
\State $M=argmin_{m\in MS}\{\sigma(C(m),CoveredEdges)|\Sigma(Cover\cup C(m))\leq\Sigma(Cover)\}$
\State \Return $C(M),M$
\EndFunction
\end{algorithmic}
\caption{GreedyOptimalCover $(G(E,V),MS)$}
\end{algorithm}
Here, $OptimalInstanceSet$ is a function that computes an optimal instance set given a motif and a set of covered edges, $\Sigma$ is the total information and $MS$ is the set of candidate motifs.  

Given a motif $m$ and a set of covered edges, finding an optimal instance set is a nontrivial optimization problem on its own. When subgraphs in the cover are not allowed to share edges, finding an optimal instance set is equivalent to finding a maximum independent set of $m$-subgraphs, that is a set of $m$-subgraphs of maximum cardinality such that no two of the subgraphs in the set have an edge in common. This problem is equivalent to the maximum independent vertex set problem  and is  NP-complete \cite{NP}. As a result some type of heuristic has to be employed. The descriptions of two such heuristics can be found in the supplemental material. Depending on the heuristic, finding an optimal instance set requires some or all of the subgraph instances of $m$ to be computed. There exist several well known algorithms that can be used for this purpose \cite{ullmann,vf2}.

\section{Empirical results}
In the following, we apply the above algorithm to several real world networks from different fields. We also consider some synthetic networks that are realizations of uniform subgraph covers with predetermined motif frequencies in order to test the heuristic. 

Due to computational resources available, the size of the subgraphs used in the analysis is limited to 5 in the directed and 6 in the undirected case. We also consider biconnected subgraph covers in relation with generalized configuration models. All results were obtained using the maximal independent set heuristic (for details see the supplemental material) for finding optimal instance sets and edge lists for encoding motifs. In the following tables, $N$  and $E$ stand for the number of vertices and edges respectively. In addition to the total information of the obtained cover $\Sigma$, the tables also show the total information of the corresponding edge cover, ERI, as a benchmark. Both these quantities are rounded to the closest integer and are given in bits.

Because of the random choices involved in finding optimal instance sets, the algorithm might find different covers for the same network on different runs. The covers shown in the figures are the best solutions obtained over multiple runs. A more detailed discussion on the variability of the heuristic can be found in the supplemental material. In Table \ref{tab:power}, the ranges of motif counts obtained over 10 runs are also shown. Here, we should stress that the proposed heuristics are primarily aimed at demonstrating the feasibility of using $\Sigma$-optimal subgraph covers as a basis of motif analysis and other heuristics might be devised for the resulting covering problem.

\subsection{Real world networks}
\begin{table}[H]
\begin{center}
	\includegraphics[width=0.48\textwidth]{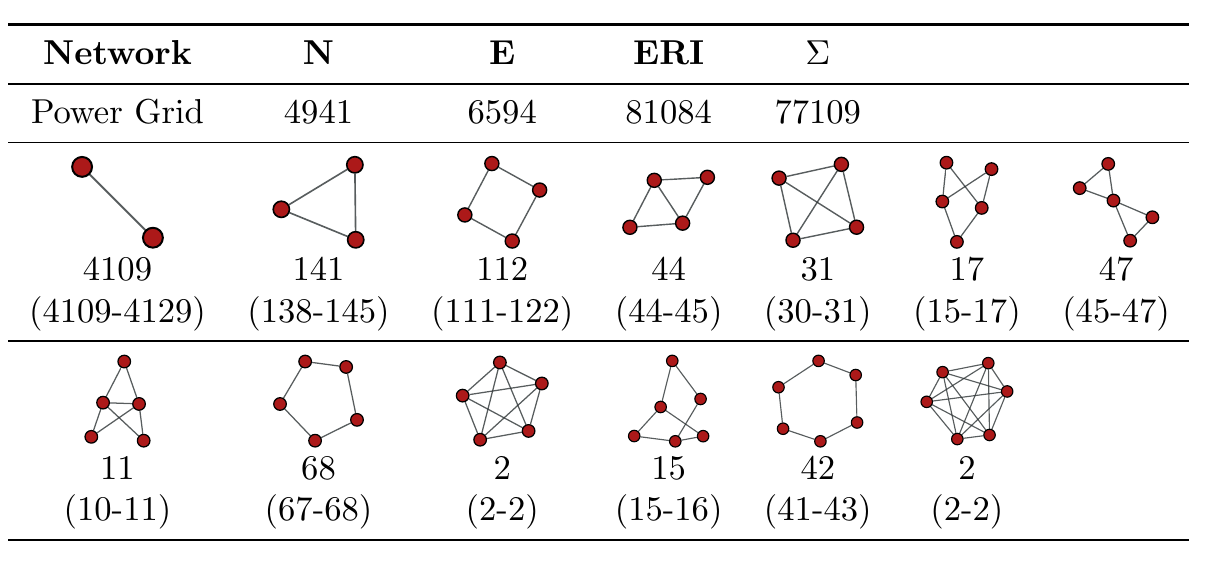}
\end{center}
\caption{The network motifs of the network representing the Western States Power Grid of the United States found using connected subgraphs up to size 6. The ranges of the motif counts of the covers found over 10 runs are also shown in parenthesis.}
\label{tab:power}
\end{table}
Table \ref{tab:power} shows the motifs found for the network representing the Western State Power Grid of the United States \cite{power}. The motif structure indicates that, among other motifs, cycles and cliques play an important role in the organization of this network. 
\begin{table}[H]
\begin{center}
	\includegraphics[width=0.48\textwidth]{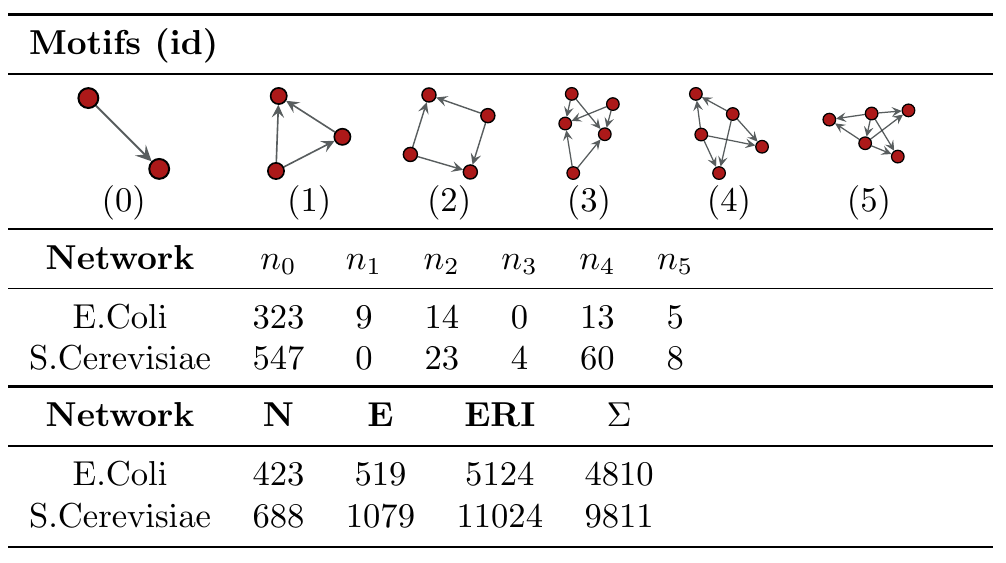}
\end{center}
\caption{The network motifs of the transcription networks of E.coli and S.cerevisiae obtained using all biconnected motifs up to size 5.}
\label{tab:gene1}
\end{table}

\begin{table}[H]
\begin{center}
	\includegraphics[width=0.48\textwidth]{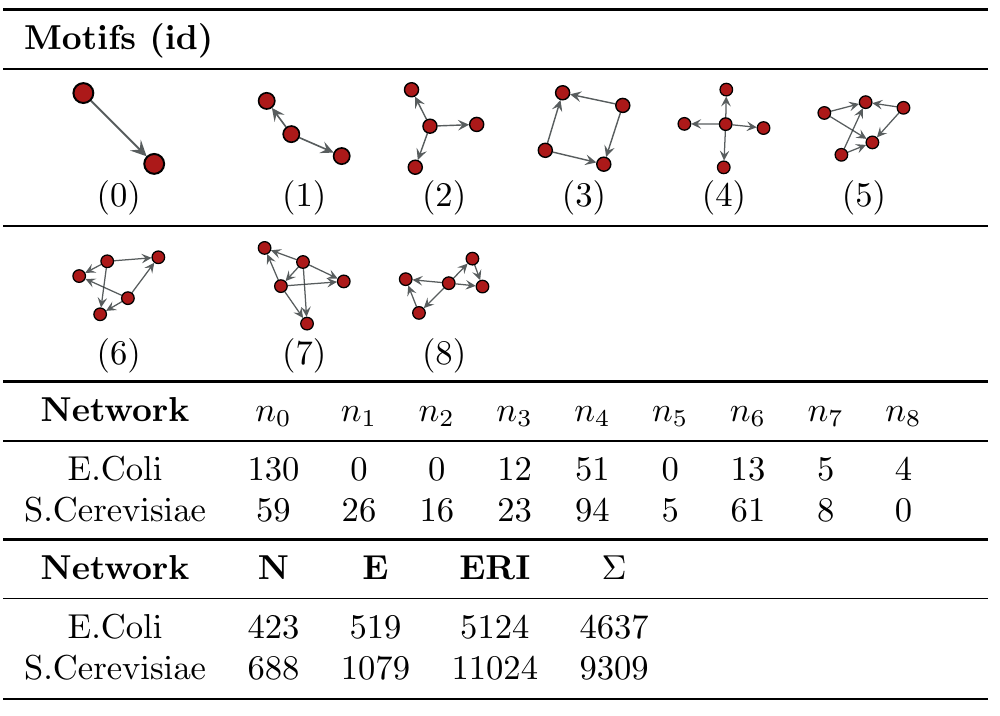}
\end{center}
\caption{The motifs of the transcription networks of E.coli and S.cerevisiae obtained using all connected motifs up to size 5.}
\label{tab:gene2}
\end{table}

In Tables \ref{tab:gene1} and \ref{tab:gene2} the motifs found for the transcription networks of E.coli \cite{ecoli} and C.cerevisiae \cite{milo} are shown. For table \ref{tab:gene1}, only biconnected motifs up to size 5 were considered while table \ref{tab:gene2} shows the results obtained using all connected subgraphs up to size 5.  Including singly connected motifs in the candidate motif set has almost no effect on the biconnected motifs and mostly results in star shaped motifs or motifs that consist of one vertex intersections of previously found biconnected motifs. The networks share 3 out of 4 motifs in the case of biconnected motifs.
\begin{table}[H]
\begin{center}
	\includegraphics[width=0.48\textwidth]{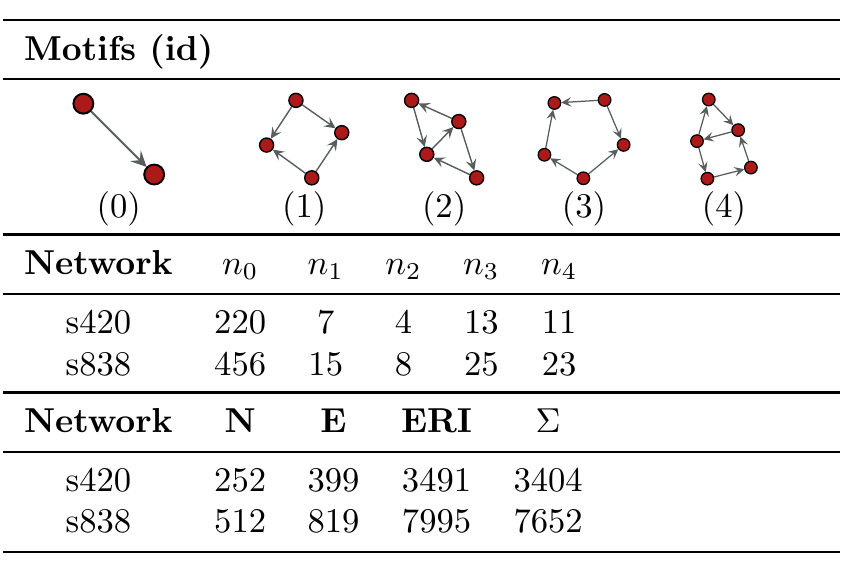}
\end{center}
\caption{The motifs of electronic circuits (digital fractional multipliers) obtained using all connected motifs up to size 5.}
\label{tab:elec}
\end{table}

In Table \ref{tab:elec} we see the results for two networks representing electronic circuits that are digital fractional multipliers \cite{milo}. For these networks the algorithm not only finds the same motifs for both networks but the motif counts in the optimal cover also scale almost exactly with network size.

\begin{table}[H]
\begin{center}
	\includegraphics[width=0.48\textwidth]{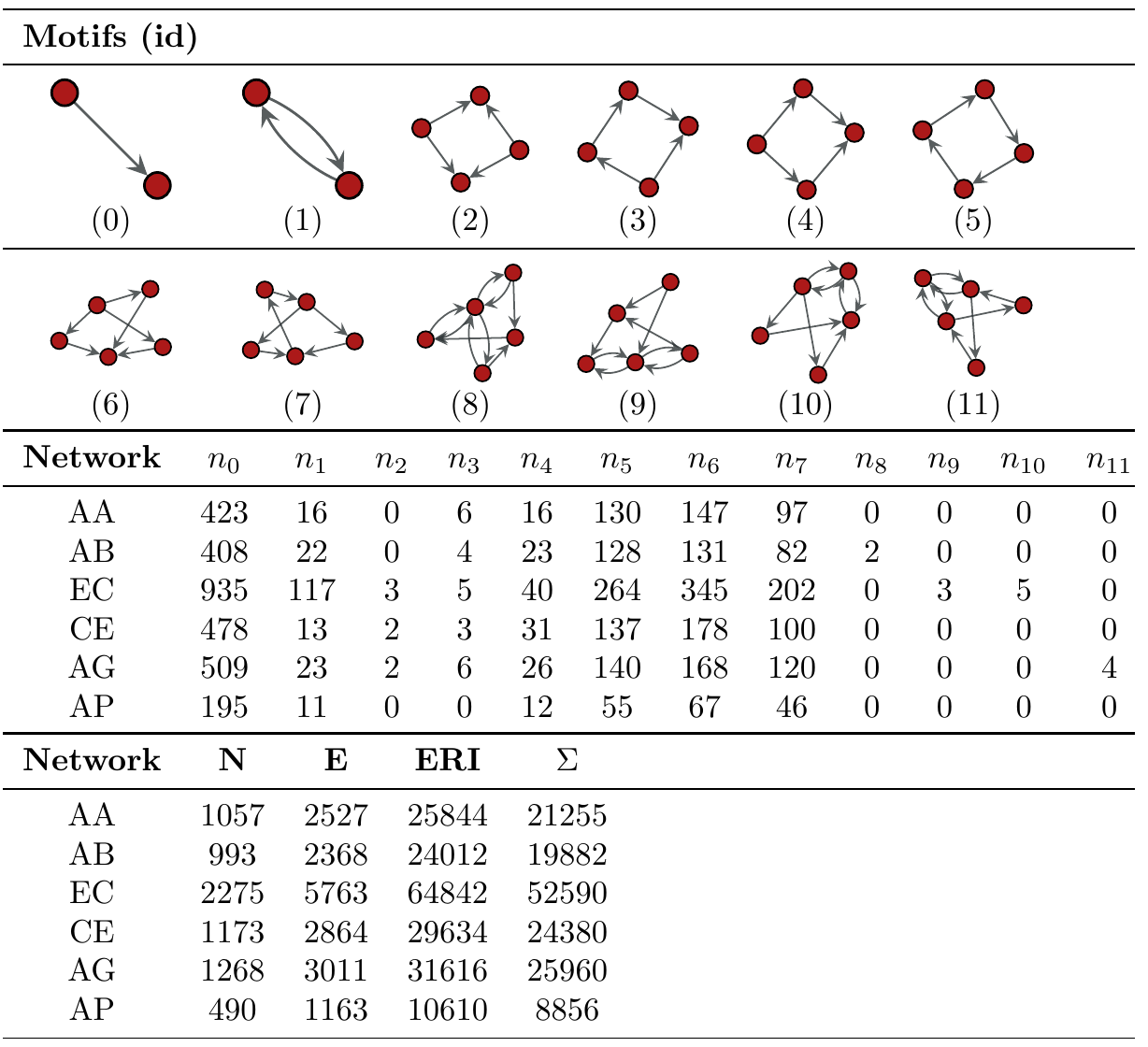}
\end{center}
\caption{The motifs of various metabolic networks  obtained using biconnected subgraphs up to size 5. AA=Aquifex aeolicus(bacteria), AB=Actinobacillus actinomycetemcomitans(bacteria),  EC=Escherichia coli (bacteria),CE=Caenorhabditis elegans (eukaryote), AG=Archaeoglobus fulgidus (archea), AP=Aeropyrum pernix(archea).}
\label{tab:metabolic}
\end{table}
In Table \ref{tab:metabolic} the network motifs found for the metabolic networks \cite{meta} of several species from different domains of life are shown. We found almost the same motifs in all of these networks and most motif counts also scale approximately with network size. 

\begin{table}[H]
\begin{center}
	\includegraphics[width=0.48\textwidth]{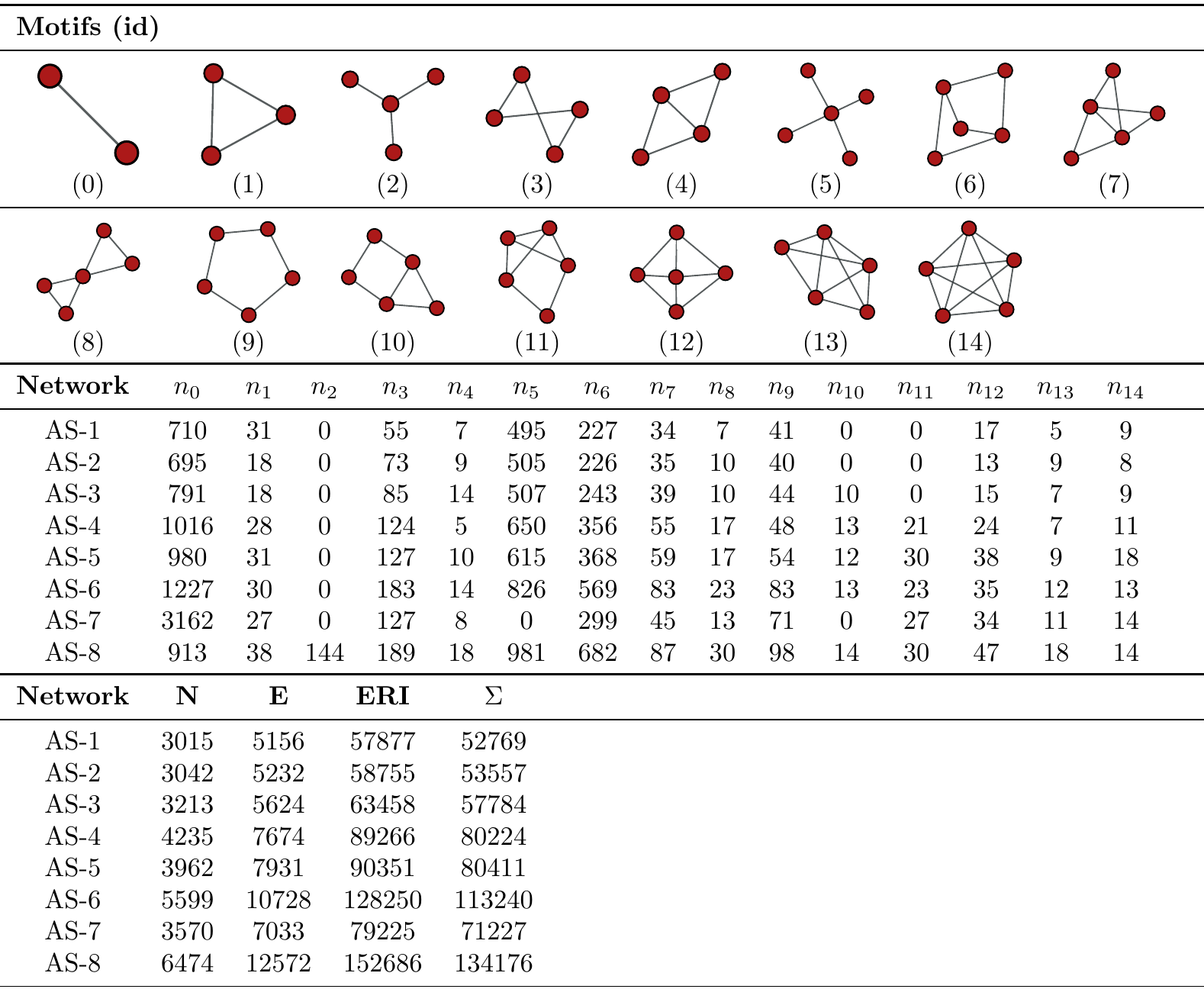}
\end{center}
\caption{The motifs of various autonomous systems networks obtained using all connected motifs up to size 5.}
\label{tab:AS}
\end{table}

Table \ref{tab:AS} shows the network motifs found in networks representing the internet at the level of autonomous systems \cite{AS}. The counts of common motifs also scale approximately with network size.

In commonly analyzed networks we find 3 and 4 node motifs that are almost identical to those found by Milo et al. \cite{milo} using a null model that conserves lower order motifs. In the transcription networks we find all the motifs found by Milo et al. though in the case S.Cerevisiae network the feed forward loop (FFL) only appears as a submotif of the larger motif consisting of 3 FFLs sharing an edge. Also for the electronic circuit networks we find the same 3 and 4 node motifs though the 3- and 4-cycles appear only as submotifs. A closer analysis shows that in these networks 3- and 4-cycles occur almost exclusively as subgraphs of larger network motifs (3-cycles: s420-19/20, s822-39/40; 4-cycles: s420-11/11, s838-23/23).

The analysis of various networks shows that networks having similar functions also have the similar motif structure. This further supports that motifs play an important role in the structural organization of complex networks. Furthermore, for networks of the same type the motif counts also scale approximately with network size. The results also show that subgraph covers can be used to obtain representations that are up to 20\% shorter compared to edge list representations. 

As previously mentioned, $\Sigma$-optimal subgraph covers can be used as a basis for associating networks with generalized configuration models which can be used to make various predictions about the properties of these networks. The method can be further tested by comparing properties of the analyzed networks with these models. However, such comparisons are beyond the scope of this article and will be treated separately in later articles. 

\subsection{Network classification}

The method can also be used to obtain a classification of networks that is similar to the classification given by Milo et al. in \cite{superfamilies}. Following the approach of \cite{superfamilies}, we use motif significance profiles that are given by the normalized c-score $\tilde{c}_m$: 

\begin{equation}
 \tilde{c}_m=\frac{c_m}{\sqrt{\sum_{m'\in M(C_\Sigma)}c_{m'}^2}}.
\end{equation} 

\begin{figure*}
\parbox{0.48\textwidth}{\includegraphics[width=0.48\textwidth]{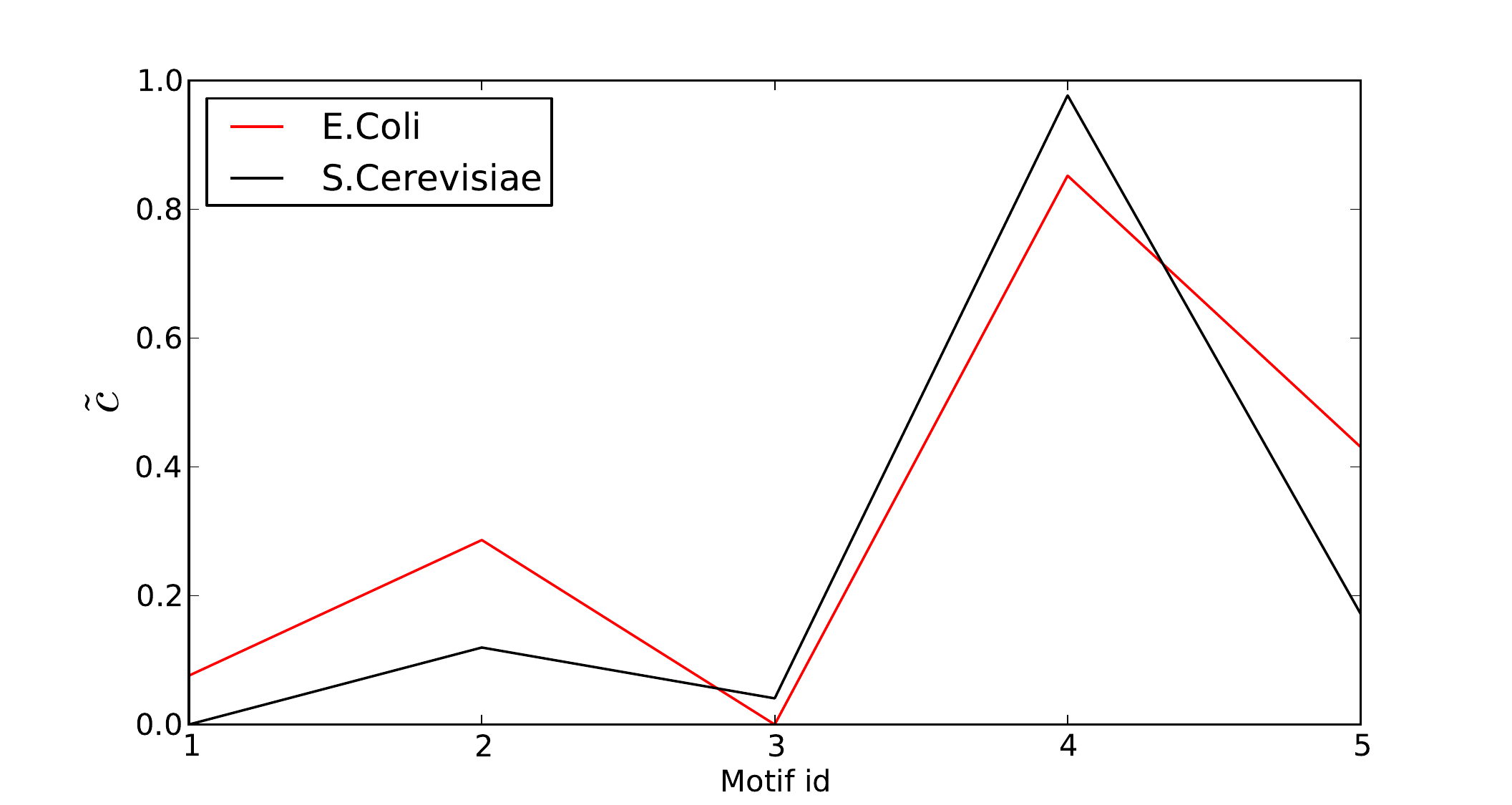}}
\parbox{0.48\textwidth}{\includegraphics[width=0.48\textwidth]{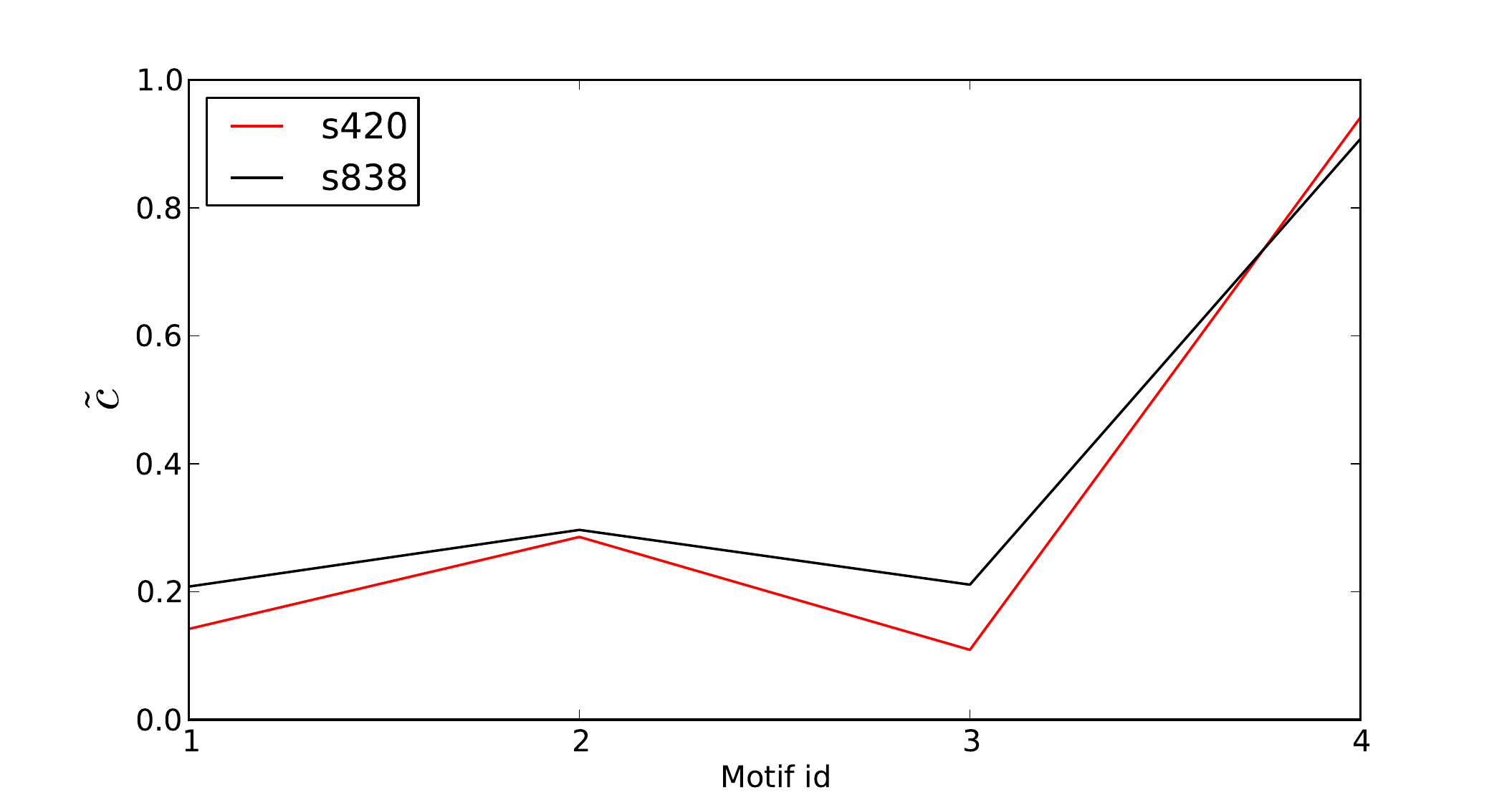}}\\
\parbox{0.48\textwidth}{\includegraphics[width=0.48\textwidth]{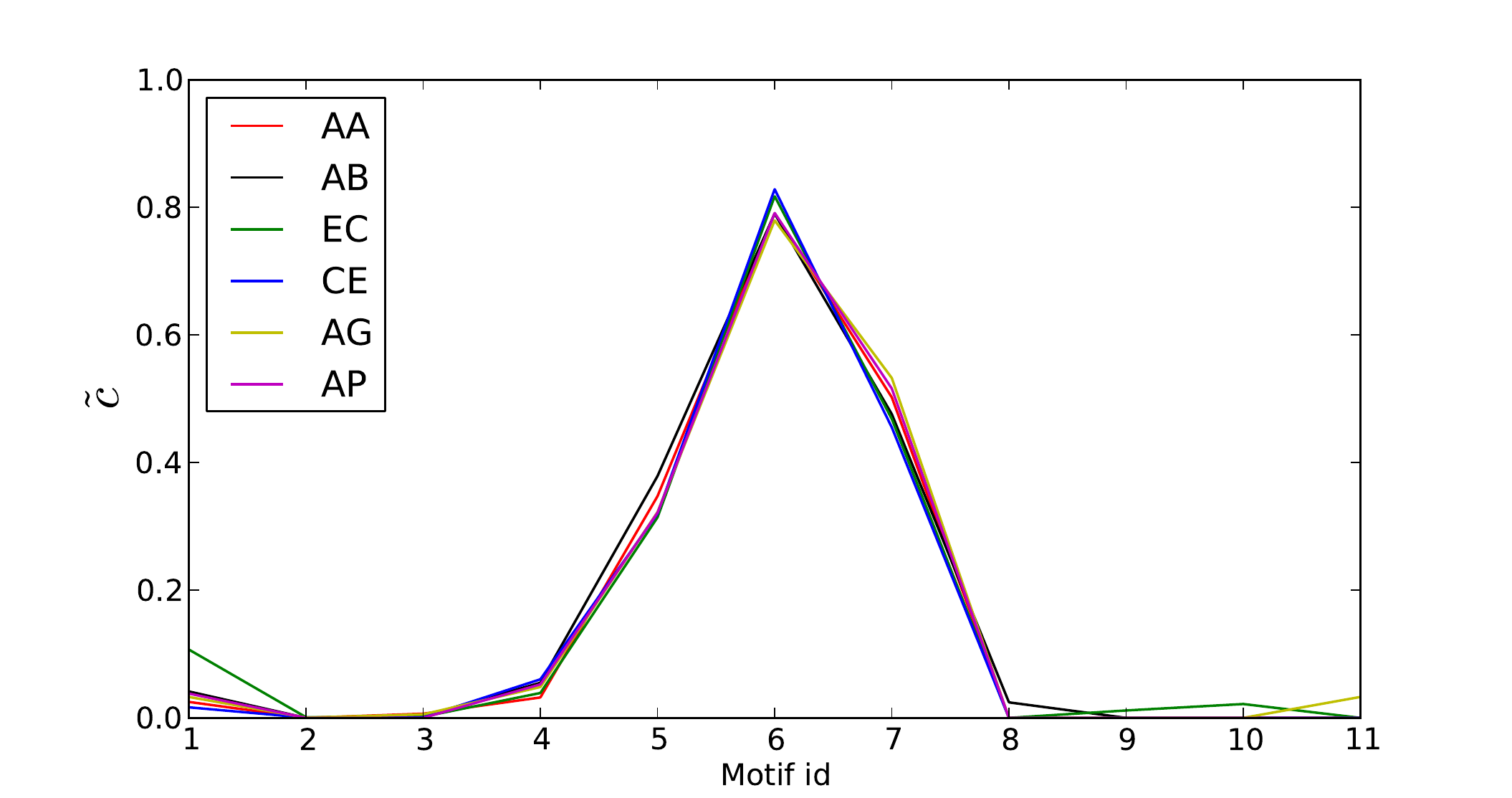}}
\parbox{0.48\textwidth}{\includegraphics[width=0.48\textwidth]{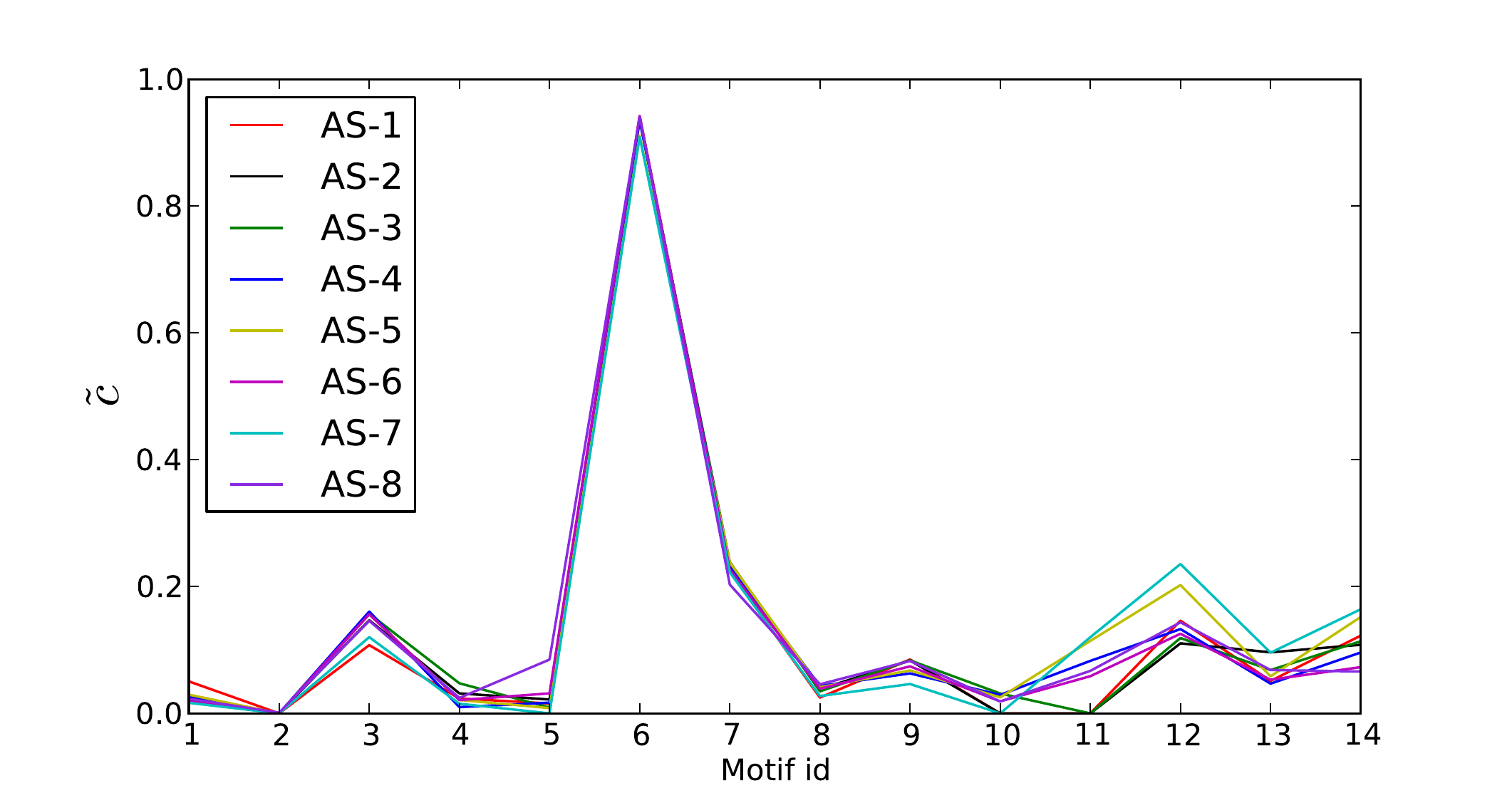}}
\caption{The motif significance profiles of various networks corresponding to the covers given in Tables II, IV, V and VI.}
\label{sigs}
\end{figure*}

The significance profiles of various network types are given in Figure \ref{sigs}. These show that networks of the same type also have very similar significance profiles. The figures only show the regions of the significance profiles corresponding to the motifs that have non-zero c-score. Compared to the full significance profile these regions are relatively small since there are 9578 connected and 7585 biconnected motifs up to size 5 in the directed case and 30 connected motifs up to size 5 in the undirected case. Thus, compared to the subgraph profiles used in \cite{superfamilies} the method provides a much finer grained classification. 

\subsection{Synthetic networks}

Finally, we also consider some synthetic networks that are realizations of uniform subgraph covers with predetermined motif counts in order to test whether the heuristic is able to recover the underlying motif set/subgraph cover in such cases. As shown in Table \ref{tab:test}, for all random networks the algorithm is able to recover the motif set. For Network 1 the algorithm recovers the underlying subgraph cover exactly. Network 2 is generated to mimic the motif structure found for an electronic circuit (s838-Table \ref{tab:elec}) and the algorithm is able to recover the original subgraph cover with only one extra subgraph. On the other hand, for Networks 3 and 4 the motif counts differ significantly from the counts of the uniform subgraph covers used to generate the networks, especially with respect to the 5-star counts. This is probably caused by the fact that these networks contain a large number of 5-stars of which only some are explicitly contained in the underlying cover. Consequently, finding an optimal instance set of 5-stars becomes more difficult. This effect is more pronounced in Network 3 because it is denser compared to Network 4.  
\begin{table}[H]
\begin{center}
	\includegraphics[width=0.48\textwidth]{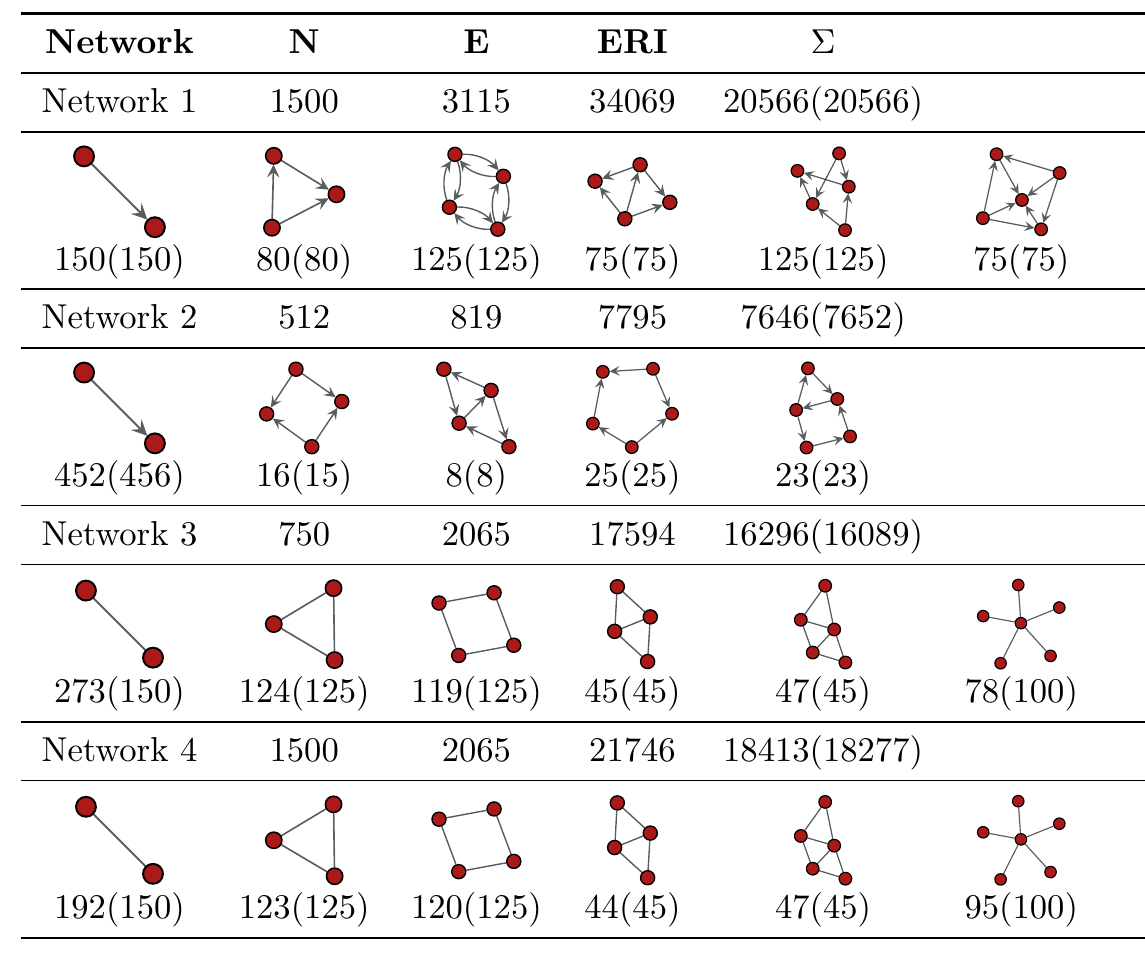}
\end{center}
\caption{The motifs obtained for networks corresponding to realizations of uniform subgraph covers. The quantities corresponding to the ensembles used to generate the networks are given in parenthesis.}
\label{tab:test}
\end{table}
 
As is the case with any heuristic, the quality of the solution depends on the structure of the network. One can construct examples where the greedy heuristic fails to recover all the motifs used to generate the network. In general, the greedy heuristic favors patterns that are dense, symmetric and occur in large numbers in the network. Thus, if the graph contains only a few copies of a motif that is not very dense, the algorithm might not be able to recover that motif. Also if a motif contains a sub-motif that is more dense and symmetric compared to the entire motif, the greedy algorithm might pick the sub-motif over the motif itself since the sub-motif covers edges more efficiently.  
 
\section{Discussion}
In this article, we introduced an alternative approach to motif analysis in networks that is based on finding a subgraph cover of the network that represents it using minimal total information. We proposed a heuristic for the resulting NP-hard optimization problem. The subgraph covers obtained for various networks show that the algorithm finds nearly identical motifs for networks with similar functions. Moreover, by considering subgraphs of various sizes simultaneously and with respect to a single global measure, the method is able to detect even large motifs consistently.

Another advantage of the method is that it provides an explicit decomposition of the network into motif subgraphs. This allows motifs to be studied within the context of the rest of the network rather than in isolation. 

We also showed that total information optimal subgraph covers can be used to match networks with random graph models that incorporate the obtained motif structure. This allows more accurate modeling of networks in general.   

Subgraph covers can readily be generalized to graphs with labeled/colored vertices and edges as well as graphs with parallel and self edges. Such labels might be chosen so that they correspond to known functional roles of vertices or the community structure of the network. On the other hand, the obtained subgraph covers could also be used as a starting point for detecting communities in networks or for inferring functional roles of vertices.  Communities that differ with respect to their internal organization can also expected to differ with respect to their motif structure. Similarly, one would expect the functional role of a vertex to be strongly correlated with the motifs it is a part of.     

The total information approach can also be extended to ensembles more general than uniform subgraph covers. Moreover, model selection approaches other than the total information approach can also be used. Such alternative formulations essentially correspond to using a different cost function in the covering problem. 

The presented analysis strongly suggests that subgraph covers can be used to compress network data. In such applications, the total information might be replaced by the expected code length of the subgraph cover. 

Finally, there is also room for improvement on the side of the heuristics. We consider this to be an important topic for further research. While the greedy algorithm can be improved, other widely used approximation schemes such as simulated annealing or genetic algorithms can also be applied to the problem.

\section{Acknowledgements}
The geng and directg tools that are a part of the Nauty \cite{nauty} package by Brendan D. McKay were used in generating the various isomorphism classes used in our analysis and the graph-tool \cite{graphtool} Python package developed by Tiago de Paula Peixoto for finding subgraphs and manipulating graphs in general. The author thanks  Prof. Jürgen Jost for constructive discussion as well as Güven Demirel and Murat Saglam for the critical reading of the manuscript. The author acknowledges the financial support of the International Max-Planck Research School at the Max-Planck Institute for Mathematics in the Sciences.  
\bibliography{motifsp}
\bibliographystyle{unsrt}

\end{document}